\documentclass[a4paper,12pt]{article}
\usepackage{amsfonts}
\usepackage{amsmath} 
\usepackage[english]{babel}
\usepackage{here}
\usepackage{multirow}
\usepackage{indentfirst}
\usepackage[dvipdfmx]{graphicx}

 \usepackage{chngcntr}

\usepackage[flushleft]{threeparttable}

\usepackage{setspace}
\usepackage[top=1.5cm,bottom=4cm,left=2.6cm,right=2.6cm]{geometry}

\title {{A New Asymmetric Copula with  Reversible Correlations and Its Application to the EU Sovereign Debt Crisis}}
\date{17 August 2021}
\author{Masahito KOBAYASHI and Jinghui CHEN \\Meiji Gakuin Unviersity and Yokohama National University\footnote{corresponding author: Masahito Kobayashi, kobayashi.masahito@gmail.com, 
 Meiji Gakuin University, Tokyo 108-8636, Japan}}

\begin{document}
\maketitle

\begin{abstract}
This paper proposes a novel asymmetric copula based upon bivariate split normal distribution. This copula can change  correlation signs of its upper and lower tails of distribution independently.
As an application, it is shown by the rolling maximum likelihood estimation that   the EU periphery countries changed sign of the lower tail correlation coefficient  from negative  to positive   after  the sovereign debt crisis started.
In contrast, Germany had negative  stock-bond correlation before and after the crisis.

\end{abstract}

\subsection*{Acknowledgment} {An earlier version of this paper was presented at the 3rd International Conference  Research in Economics \& Social Science in Shiga University, Hikone, Japan, on November 28-29, 2019. The authors would like to thank Kentaro Iwatsubo, Yoshihiro Kitamura, Michael McAleer, Daisuke Nagakura, Craig Parsons, Kenji Wada, and Clinton Watkins for their helpful suggestions and the seminar participants at Keio University, Kobe University, and Yokohama National University for their valuable comments. This work was supported by JSPS KAKENHI JP 15K03394.

Keywords: Copula, Asymmetry, Rolling Estimation

\section{Introduction} 

Recently, Yoshiba (2013) revealed that  stock and bond returns correlation had sign changes  and asymmetry during  financial crisis using nonparametric copula: the correlation coefficient changed from negative to positive  and the correlation was stronger on the downside than on the upside. This co-occurrence of correlation asymmetry and sign-reversal of financial asset returns during  financial crisis has not been reported by researchers, though they have been independently well-studied issues.

Many authors such as Patton (2006) and Okimoto (2008) analyzed asymmetry of financial returns  by copula,  which was first proposed by Sklar (1959).   It  is a joint distribution function with  uniform marginal distribution.   By transforming the marginal variables of the copula  it  can  construct a  joint distribution function with  the same dependence structure as that of the copula and   marginal distribution specified by univariate models such as GARCH . 
The  parametric copulas, especially  the Joe-Clayton copula, is useful in that  their parameters are directly  related to  tail dependence and hence their parameter estimates can be interpreted from the viewpoint of economists with ease.  

Correlation reversal in  financial crisis has been  reported by Ohmi and Okimoto (2016) and Dufour et al. (2017), among others.
However, the  co-occurrence of asymmetry and correlation reversal of financial returns in  financial crisis cannot be analyzed by  parametric  bivariate copulas, such as the Joe-Clayton copula, because they always have a positive correlation coefficient as shown by Li and Kang(2018).

Recently, several authors  analyzed financial  returns  using mixture copula ( Cai and Wang, 2014; Yang et al., 2018, Liu et al. 2019),  which  is a weighted sum of parametric copulas.  However, up to our knowledge, no parametric family of  mixture copulas that expresses  asymmetry and correlation reversal has been proposed.

In this paper we propose a novel two-parameter asymmetric copula, whose correlation signs in the upper and lower tail areas can change independently. We construct this copula  using the inversion method from bivariate split normal distribution consisting of two bivariate half normal density functions with different correlation coefficients connected on the negative 45 degree line. This distribution is a special case of  the general bivariate split normal distribution , which was first proposed by Geweke (1989) and investigated by Villani and Larson (2006) in detail. We estimate the  lower  and upper tail correlations of unobservable split normal distribution  that underlies the copula from  the  dependence structure  of the copula by the maximum likelihood method(MLE).

This copula was applied to  stock and bond returns of Germany and the five EU periphery countries, namely, Greece, Ireland, Italy, Portugal, and Spain and the parameters were estimated using the rolling maximum likelihood method. It was found that, in the early stage of the crisis, the stock-bond distribution of the EU periphery countries had  positive lower tail correlation and negative or near-zero upper tail correlation. This strong asymmetry  implies  that the price of one asset  tends to decrease when the price of the other asset decreases  and that, when the price of one asset increases, the price of the other asset does not always increase. This suggests capital flight from these countries.  In contrast, Germany had no correlation reversal of stock-bond returns.

The rest of the paper is organized as follows. In Section 2  the  new copula is derived.  In Section 3 the performance of MLE is examined using Monte Carlo simulation. Lastly, in Section 4, results of the empirical analysis are presented.

\section{Model and Distribution}

\subsection{Split normal distribution}

We first define  bivariate split normal distribution, upon which our new copula is based.
It is  constructed by two bivariate normal density functions with different correlation coefficients continuously connected on the negative 45 degree line by height adjustment. This distribution is a special case of the original split normal distribution, which was first proposed by Geweke (1989) and investigated by Villani and Larson (2006);  their original bivariate split normal distribution  are split  along two axes, whereas our split normal distribution is split only along negative 45 degree line and  is symmetric with respect to the  positive 45 degree line. This copula is  exchangeable in the terminology of Nelson (2006, p.38).

This assumption  of symmetry is useful in that  the lower and upper tail correlation  coefficients summarize the   asymmetry 
between upper right and bottom-left tails of the distribution, namely the asymmetry  of  the distribution of two financial returns when they are  both positive or both negative. This asymmetry   is an important research topic  in finance and  has been studied conventionally  by parametric exchangeable copulas, such as the Joe-Clyaton and Gunbel copulas.   
It is an interesting problem to apply copula functions generated from general split normal distribution   with two axes along positive and negative 45 degree lines  and to consider its implications, but is left to further research, since we are interested only in the behavior of upper and lower tail of the return distribution. 

In this paper we define a bivariate split normal density defined as follows:
\begin{equation}\label{f}\textrm{Split normal density: }
f_{W,V}(w,v)=\left\{\begin{array} {cc}
a_U\times \phi(w,v,\rho_U, \sigma^2_U)\quad \textrm{if}\quad w+v> 0,
\\
a_L\times \phi(w,v,\rho_L, \sigma^2_L)\quad \textrm{if}\quad w +v\le 0,
 \end{array}\right. 
\end{equation}
where $\phi$ is the symmetric bivariate normal density with correlation $\rho$, zero mean, and variance $\sigma^2$ defined by
\begin{equation}\phi(w,v,\rho,\sigma^2)=\frac{1}{2 \pi \sigma^2 \sqrt{1-\rho^2}}
\exp\left(-\frac{w^2-2\rho wv+v^2} {2(1-\rho^2)\sigma^2} \right).
\end{equation}
Hereafter,  the area $w+v > 0$ is referred to as the upper tail area and the area $w+v\le  0$  as the lower tail area, where there is no fear of ambiguity.

The weight constants $a_L$ and $a_U$ should satisfy 
\begin{eqnarray}\label{cond1}
a_U+a_L=2
\end{eqnarray}
from the condition that  $f_{W,V}(w,v)$ in (\ref{f}) is a probability density function.

 The heights of the two parts of the density functions are matched on the boundary $w+v=0$  to connect the two parts continuously. This identity condition, which  is expressed  as 
\begin{eqnarray*}
\frac{ a_U}{ \sigma_U^2 \sqrt{1-\rho_U^2}}
\exp\left[\frac{-w^2} {(1-\rho_U)\sigma_U^2} \right]
\equiv \frac{ a_L}{\sigma_L^2 \sqrt{1-\rho_L^2}}
\exp\left[\frac{-w^2} {(1-\rho_L)\sigma_L^2} \right] ,\, -\infty < w < \infty,
\end{eqnarray*}
is reduced to the following two equations:
\begin{eqnarray}
\label{cond2}
a_L/a_U= \frac{ \sigma_L^2\sqrt{1-\rho_L^2}}  { \sigma_U^2 \sqrt{1-\rho_U^2}},\quad
\end{eqnarray}
\begin{eqnarray}
\label{cond3}
\sigma_U^2= \sigma_L^2(1-\rho_L)/(1-\rho_U ).
\end{eqnarray}

Then the split normal distribution defined by (\ref{f}) is uniquely determined by the parameters $(\sigma_L, \rho_U, \rho_L) $, since $a_U$ and $a_L$ are uniquely determined by $( \sigma_U, \sigma_L, \rho_U, \rho_L )$ using equations (\ref{cond1}) and (\ref{cond2}), and $\sigma^2_U$ is determined by  $(\sigma_L, \rho_U, \rho_L) $ using (\ref{cond3}).
 
 Figure \ref{fig:split_normal} illustrates the contours of  split normal densities when the upper and lower correlations are (-0.6, -0.2) and (-0.85, 0.9).
 \begin{figure}[H]
  \begin{center}
  \includegraphics{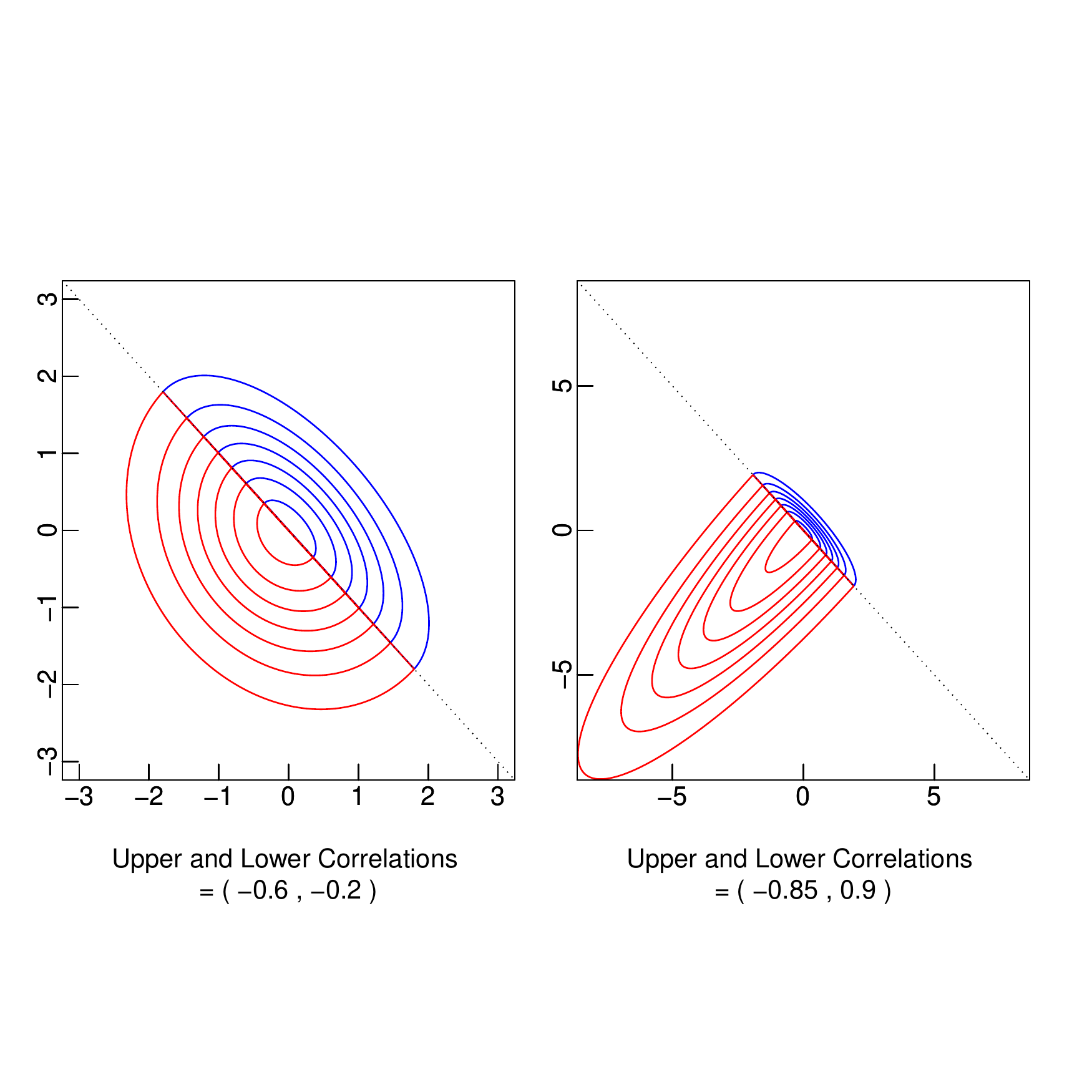}

  \caption{Contour plots of split normal densities.}   
 \label{fig:split_normal}
  \end{center}

\end{figure}
\subsection{Split Normal Copula}

We here construct  a new copula  from the split normal distribution in order to express asymmetric  dependence structure that can change sign. 
Let  $W$ and $V$ denote unobservable continuous random variables that follow bivariate normal distribution  defined by (\ref{f}) to construct . 
Copula, which was first proposed by Sklar (1959), is a joint distribution function with  uniform marginal distribution and provides a flexible  tool to construct correlation structure independently of marginal distribution. 
Bivariate split normal copula is  defined by
\begin{equation}\label{copula}
\textrm{Copula}: C(x,y) = F(F_W^{-1}(x), F_V^{-1}(y)) ,\quad 0<x<1,\quad 0<y<1,
\end{equation}
 from the  distribution function $$F(w,v)=\int ^w_{-\infty} \int ^v_{-\infty}f_{W,V}(v_1,v_2)dv_1dv_2,$$ where $F_W(w)$ is the marginal distribution function, and $F_W^{-1}(x)$ is its inverse. 
Then   $X = F_W(W)$ and $ Y = F_V(V) $ follow  uniform distribution  with joint distribution function $$ F(F_W^{-1}(x), F_V^{-1}(y)) =P(X<x, Y<y).$$
  Then the density function of the uniformized variables $X = F_W(W)$ and $ Y= F_V(V) $ is expressed as
\begin{equation}\label{copulad1}
\textrm{Copula Density Function: }c(x,y) = \frac{f_{W,V}(F_W^{-1}(x), F_V^{-1}(y))}{f_W(F_W^{-1}(x))f_V(F_V^{-1}(y))} .
\end{equation}
The merit of copula is that   monotonic transformation, such as $W = F_W^{-1}(X)$ and $V = F_V^{-1}(Y)$, does not change  quantile dependence, defined by
$P(V > F_V^{-1}(p)| W > F_W^{-1}(p))$  for $0<p<1$. This result can be easily shown, 
noting that  $$P(V > F_V^{-1}(p) | W > F_W^{-1}(p))=P(F_V(V )> p | F_W(W) > p) = P(X> p | Y > p)$$
 and that $p=F_Y^{-1}(p)$  and $p=F_X^{-1}(p)$.  This is referred to as the copula invariance property (Nelson, 2006, p.25).
 Then It follows that any join distribution function that has the same copula has the same  correlation structure defined by the  quantile dependence.

It should be noted that  we can set  \begin{equation}
\sigma_L = 1
\end{equation}
 for identification using  the copula invariance property, since,   
as stated above,  the same copula is derived from any strictly monotonic transformations of the marginal random variables and hence  
the copula function is unchanged  when the  random variables $W$ and $V$ in (\ref{f})  is divided  by $\sigma_L $.
Then  the  split normal copula   can be parameterized by $(\rho_L, \rho_U)$.

Figure \ref{fig:split_normal_copula} illustrates the contours of 
split normal copula distribution  where the marginal variables are transformed so as to have normal marginal distribution; they are  
derived from the  split normal distribution in Figure \ref{fig:split_normal} when the upper and lower correlations are (-0.6, -0.2) and (-0.85, 0.9).

\begin{figure}[H]
  \begin{center}
  \includegraphics{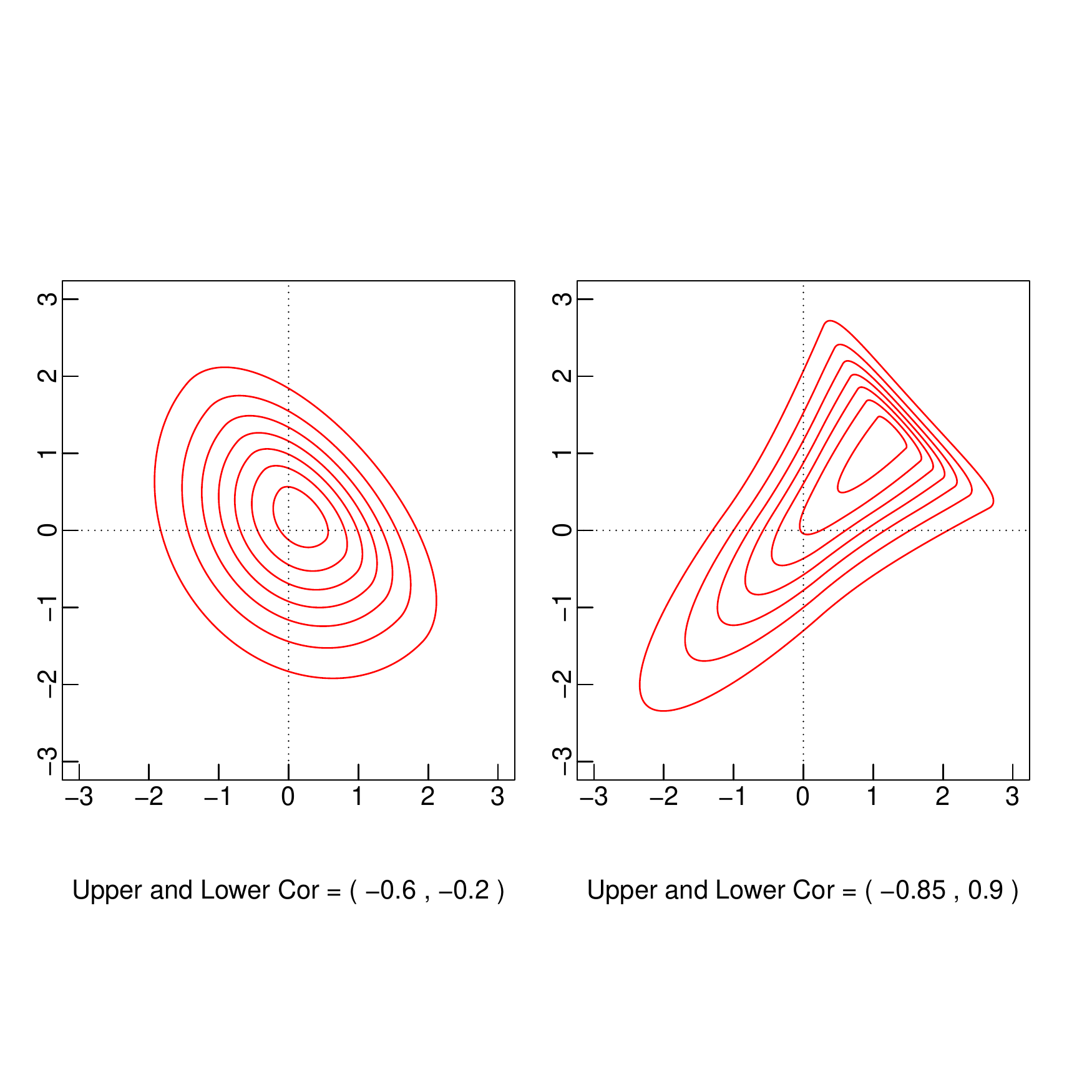}
  \caption{Contour plots of split normal copula where  marginal variables are  transformed so as to have normal distribution, for illustration.  They are derived from split normal distributions in Figure 1.}   
 \label{fig:split_normal_copula}
  \end{center}

\end{figure}

In practice, it is  assumed that  the uniformized variables $X$ and $Y$ that follow split normal copula distribution are observed, or estimated using univariate models of $X$ and $Y$, whereas the background variables $W$ and $V$ that follow split normal distribution are unobservable. We will estimate $\rho_U$ and $\rho_L$, namely the parameters of the distribution of  $W$ and $V$, from the observed values of  $X$ and $Y$ by maximizing the likelihood function defined below.

%
%
%
%

\subsection{Numerical Evaluation of the Likelihood}
The copula density (\ref{copulad1}) for the split normal distribution in (\ref{f}) has no analytical expression. Then, given the joint distribution function of $W$ and $V$ that follow split normal distribution function (\ref{f}), we can evaluate $F_W(w)$, $f_W(w)=F'_W(w)$, and $F_W^{-1}(x)$ numerically as shown below. 
We have $F_V(\cdot)=F_W(\cdot)$, $F_V^{-1}(\cdot)=F_W^{-1}(\cdot)$, and $f_{V}(\cdot)=f_W(\cdot)$ from the assumed symmetry of $f_{W,V}(w,v)$. Then substituting $F_V^{-1}(\cdot)$ and $F_W^{-1}(\cdot)$ into (\ref{copulad1}), we have copula density function. 

First, decompose $F_W(w)$ as
\begin{eqnarray}\label{cdf}
F_W(w)
=P(W<w,0<W+V)+ P(W<w,0>W+v),
\end{eqnarray} 
where $(W, V)$ in the first term on the right-hand side lies in the upper tail area of the density function (\ref{f}), and $(W,V)$ in the second term on the right-hand side lies in the lower tail area of the density function (\ref{f}).

Then we have
\begin{equation}\label{cdf1}
P(W<w, W+V>0) = a_U\times P(Z_1<w,Z_1+Z_2>0 ),\end{equation}
where $Z_1$ and $Z_2$ are normally distributed random variables with zero means,  variances $\sigma_L^2$ and $\sigma_U^2$, and  correlation $\rho$.
Analogously, we have %
\begin{equation}\label{cdf2}
P(W<w, W+V<0) = a_L\times P(Z_1<w, 0>Z_1+Z_2).
\end{equation}
 We then calculate
$F_W(w)$ numerically  at discrete values $w_1<\,\,\ldots < w_M$ combining the results of  (\ref{cdf1}) and (\ref{cdf2}) in (\ref{cdf}). In this paper we use $M = 50$.
Then, $F_W(w)$ as a smooth function of $w$ is derived by applying spline interpolation to $(w_i, F_W(w_i)),\,\, i = 1,\ldots, M$.  The density function $f_W(w)$ is obtained as a  derivative of $F_W(w)$.
The quantile function $F_W^{-1}(\cdot)$, as a continuous function, is obtained analogously from $( F_W(w_i),w_i)$ using spline interpolation.

\section{Monte Carlo Experiments}\subsection{Distribution of Estimators}
We examine the performance of estimators of the  upper  and lower tail correlation coefficients $\rho_U$ and $\rho_L$ in (\ref{f}) by  grid search maximum likelihood method using Monte Carlo experiment.
Table \ref{mean}  shows the sample mean, standard deviation (SD), skewness (SKEW), and kurtosis (KURT) of the estimates of upper tail correlation coefficient for a sample size of 100  with 500 iterations. The summary statistics for lower tail correlation coefficient can be obtained from these tables using the symmetry between $\rho_L$ and $\rho_U$ and hence omitted.  We used grid width   0.02, which is wider than that used in the empirical analysis. 

Table \ref{mean} shows that the distribution of the estimator is strongly skewed toward the origin when the correlation coefficient to be estimated is large in absolute value. Thus, normal approximation cannot be used in hypothesis testing when the sample size is 100. 

\subsection{Testing for  Zero Correlation}
Here, we obtain the critical values of the one-sided test  for the null hypothesis of zero upper tail correlation by Monte Carlo simulation. Artificial data are generated under the null hypothesis that the upper tail correlation is zero whereas the lower tail correlation is  nonzero. 
Table \ref{cv} presents 5, 10, 90, and 95 percentiles of the null distribution of the estimated upper tail correlation coefficient, which shows the critical values of the one-sided test at five and ten percent levels of significance.
 We calculated critical values  when the value of correlation coefficients of the upper tail is  -0.8, -0.6, -0.4, -0.2, 0.0, 0.2, 0.4, 0.6, and 0.8. In practice, critical values in other cases can be easily obtained by linear interpolation.

The percentiles of the estimated correlation coefficients  $\rho_U$ and $\rho_L$  for sample size 100 were calculated  using Monte Carlo simulation with 500 iterations. MLE were obtained using grid search with grid width  0.02. It shows that the acceptance region widens as correlation coefficient of the other tail increases.

\begin{table}

\begin{center}
\caption{Moments of MLE for LowerTail Correlation Coefficient }

\small
\label{mean}
\begin{tabular}{cccccccccccccccccccccccccccccccc}
\multicolumn{2}{c}{True Cor}&\multicolumn{4}{c}{Estimated LowerTail Cor }&&\multicolumn{2}{c}{True Cor}&\multicolumn{4}{c}{Estimated LowerTail Cor}\\
$\rho_U$	&$\rho_L$	&	MEAN	&	SD&	SKEW	&KURT			&& $\rho_U$	&$\rho_L$	&	MEAN	&	SD&	SKEW	&KURT	\\
\hline

  0.6 & 0.6 & 0.57 &0.16 &-2.72 &   18.13  && -0.2 & 0.6 & 0.59 &0.10 &-1.16 &    3.43\\ 
  0.6 & 0.4 & 0.37 &0.22 &-1.10 &    1.36 &&  -0.2 & 0.4 & 0.39 &0.14 &-0.75 &    1.45\\ 
  0.6 & 0.2 & 0.15 &0.28 &-1.02 &    1.89 &&  -0.2 & 0.2 & 0.17 &0.18 &-0.59 &    0.27\\ 
  0.6 & 0.0 &-0.05 &0.31 &-0.47 &    0.06 &&  -0.2 & 0.0 & 0.00 &0.19 &-0.47 &    0.32\\ 
  0.6 &-0.2 &-0.25 &0.30 &-0.07 &   -0.23 && -0.2 &-0.2 &-0.21 &0.21 &-0.17 &    0.09\\ 
  0.6 &-0.4 &-0.43 &0.28 & 0.20 &   -0.45 &&  -0.2 &-0.4 &-0.41 &0.19 & 0.23 &    0.32\\ 
  0.6 &-0.6 &-0.61 &0.23 & 0.73 &    0.75 &&  -0.2 &-0.6 &-0.60 &0.15 & 0.09 &   -0.10\\ 
  0.4 & 0.6 & 0.58 &0.13 &-1.29 &    3.64 &&  -0.4 & 0.6 & 0.60 &0.09 &-0.95 &    1.79\\ 
  0.4 & 0.4 & 0.36 &0.21 &-1.45 &    3.52 &&  -0.4 & 0.4 & 0.39 &0.13 &-0.65 &    0.76\\ 
  0.4 & 0.2 & 0.16 &0.23 &-0.89 &    1.02 &&  -0.4 & 0.2 & 0.18 &0.16 &-0.70 &    1.43\\ 
  0.4 & 0.0 &-0.06 &0.28 &-0.61 &    0.51 &&  -0.4 & 0.0 & 0.00 &0.18 &-0.23 &    0.32\\ 
  0.4 &-0.2 &-0.24 &0.26 &-0.25 &   -0.21 &&  -0.4 &-0.2 &-0.21 &0.18 & 0.07 &   -0.01\\ 
  0.4 &-0.4 &-0.44 &0.25 & 0.25 &   -0.32 &&  -0.4 &-0.4 &-0.41 &0.17 & 0.25 &    0.36\\ 
  0.4 &-0.6 &-0.61 &0.20 & 0.29 &   -0.20 &&  -0.4 &-0.6 &-0.60 &0.14 & 0.24 &    0.10\\ 
  0.2 & 0.6 & 0.58 &0.13 &-2.67 &   17.52 &&  -0.6 & 0.6 & 0.59 &0.08 &-0.55 &    0.43\\ 
  0.2 & 0.4 & 0.38 &0.18 &-1.06 &    2.40 &&  -0.6 & 0.4 & 0.40 &0.12 &-0.48 &    0.44\\ 
  0.2 & 0.2 & 0.17 &0.22 &-0.99 &    1.86 &&  -0.6 & 0.2 & 0.19 &0.15 &-0.68 &    2.16\\ 
  0.2 & 0.0 &-0.04 &0.24 &-0.52 &    0.66 &&  -0.6 & 0.0 & 0.00 &0.16 &-0.38 &    0.63\\ 
  0.2 &-0.2 &-0.21 &0.24 & 0.05 &   -0.22 &&  -0.6 &-0.2 &-0.20 &0.15 &-0.04 &    0.80\\ 
  0.2 &-0.4 &-0.42 &0.22 & 0.31 &    0.14 &&  -0.6 &-0.4 &-0.40 &0.15 & 0.14 &    0.40\\ 
  0.2 &-0.6 &-0.61 &0.18 & 0.39 &    0.18 &&  -0.6 &-0.6 &-0.61 &0.12 & 0.15 &    0.11\\
  0.0 & 0.6 & 0.59 &0.10 &-0.84 &    1.80\\ 
  0.0 & 0.4 & 0.39 &0.15 &-0.81 &    1.59\\ 
  0.0 & 0.2 & 0.17 &0.20 &-0.79 &    1.67\\ 
  0.0 & 0.0 &-0.02 &0.21 &-0.37 &    0.21\\ 
  0.0 &-0.2 &-0.23 &0.21 &-0.06 &    0.15\\ 
  0.0 &-0.4 &-0.42 &0.21 & 0.13 &   -0.07\\ 
  0.0 &-0.6 &-0.62 &0.15 & 0.68 &    0.85\\ 
 \hline
\end{tabular}
\begin{flushleft}
Note:  Standard deviation, skewness , and kurtosis are denoted by  SD, SKEW, and KURT, respectively.  Artificial data are generated using the upper and lower tail correlation coefficients  (denoted by Cor for short) given in the first and second columns.  Moments are estimated for sample size 500 by Monte Carlo simulation with 100 iterations.
\end{flushleft}

\end{center}

\end{table}

\begin{table}

\begin{center}
\caption{Percentiles of the Estimated LowerTail Correlation Coefficient $\rho_L$ When its True Value   Zero
 }
\begin{tabular}{ccccccccccccccc}
\\
\\True $\rho_U$&\multicolumn{4}{c}{Percentiles} \\
  	&	 0.05 		&		0.10&	0.90&	0.95\\\hline
0.8		&	 -0.84		&		-0.62&	0.38 &	0.50\\
0.6		&	 -0.58		&		-0.41&	0.31 &	0.41\\
0.4		&	 -0.52		&		-0.41&	0.29 &	0.37\\
0.2		&	 -0.50		&		-0.38&	0.27 &	0.35\\
0		&	 -0.35		&		-0.29&	0.25 &	0.31\\
-0.2	&	 -0.41		&		-0.29&	0.27 &	0.31\\
-0.4	&	 -0.31		&		-0.21&	0.23 &	0.29\\
-0.6	&	 -0.29		&		-0.23&	0.19 &	0.25\\
-0.8	&	 -0.23		&		-0.17&	0.19 &	0.23\\
\hline
\\
\end{tabular}
\label{cv}
\begin{minipage}{16.1cm}

Note: The percentiles  were calculated using  with 1000 iterations. The sample distribution were artificially generated under the assumption that the true lower tail correlation is zero. 

\end{minipage}
\end{center}

\end{table}

\newpage
\section{Empirical Analysis}

\subsection{Stock-Bond Correlation}

Co-occurrence of  sign-changes and asymmetry of stock-bond correlations has not been much discussed in previous studies, though they have been independently important issues in financial econometrics.

Correlation reversal of stock-bond returns in the EU crisis was reported by Dufour et al. (2017) and Ohmi and Okimoto (2016).
They found that the stock-bond correlations changed signs from negative to positive during the financial crisis and remained positive even after the crisis ended.

Copulas have been extensively used to analyze correlation asymmetry between financial returns. 
Patton (2006) estimated  two-parameter Joe-Clayton copula  in a time series setting. Okimoto (2008) used Markov switching model and copula to analyze international equity markets. Christoffersen et al. (2012) estimated dynamic asymmetric correlations in large cross-sections, generalizing the dynamic conditional correlation model of Engle (2002).

Up to our knowledge, only Yoshiba (2013) reported that the stock-bond correlations had sign changes and correlation asymmetry at the same time during the financial crisis. He used a nonparametric copula because conventional parametric copulas cannot express asymmetry and sign changes of stock-bond correlations. For example, the Joe-Clayton copula always has a positive correlation (Li and Kang, 2018) and  cannot express upper tail and lower tail correlations with different signs. Then the conventional   parametric copulas cannot analyze stock-bond correlation during the EU sovereign debt crisis. 
%

In this paper we employ the inversion method (Nelson, 2006, p.51) to construct a two-parameter asymmetric copula based on split normal distribution, whose upper and lower tail correlation coefficients can change values and signs independently. We also analyze  asymmetric and reversible correlation  of  the stock-bond returns in the EU crisis.

\subsection{Data}

\begin{table}

\begin{center}
\caption{Summary Statistics of Financial Returns}
\begin{tabular}{lllllllll}Country& Assets &MEAN&SD &MIN&MAX&SKEW&KURT&$\tau$
\\ \hline \hline
  
\label{summary}
  
\multirow{2}{*}{Greece }&  
     FTSE ATHEX 20 
     
    &	-0.357	&	5.185	&	-23.160	&	18.650	&	-0.421	&	1.927	&

     \multirow{2}{*}{0.292}
\\ 
& 10 Yr Gov B

&	-0.008	&	9.131	&	-61.953	&	161.233	&	8.250	&	158.619	&

\\
\hline
\multirow{2}{*}{Ireland} & FTSE ISEQ ALL 

&	-0.028	&	4.031	&	-37.103	&	16.616	&	-1.642	&	13.827	&

			\multirow{2}{*}{0.055}
 \\ 
& 10 Yr Gov B

&	0.043	&	2.250	&	-15.061	&	21.381	&	0.573	&	21.061	&

\\ \hline
\multirow{2}{*}{Italy}&  FTSE MIB 

&	-0.074	&	3.447	&	-24.360	&	10.472	&	-1.135	&	5.279	&

\multirow{2}{*}{0.182}
\\ 
& 10 Yr Gov B

&	0.025	&	1.359	&	-5.824	&	10.447	&	0.832	&	8.306	&

\\ \hline
\multirow{2}{*}{Portugal} & PSI-ALL Share 

&	0.028	&	2.764	&	-20.528	&	7.677	&	-1.202	&	5.906	&

\multirow{2}{*}{0.204}

\\ 
& 10 Yr Gov B

&	0.024	&	2.974	&	-18.871	&	16.597	&	-0.074	&	10.159	&

\\ \hline
\multirow{2}{*}{Spain}& IBEX 35

&	-0.009	&	3.336	&	-23.827	&	11.101	&	-0.924	&	4.942	&

  \multirow{2}{*}{0.153}

\\ 
& 10 Yr Gov B

&	0.028	&	1.606	&	-6.545	&	10.521	&	1.051	&	9.066	&

\\ \hline 
\multirow{2}{*}{Germany}& DAX 

&	0.141	&	3.094	&	-24.347	&	14.942	&	-1.012	&	8.100	&

\multirow{2}{*}{-0.273}

\\ 
& 10 Yr Gov B
&	0.046	&	0.995	&	-3.606	&	3.676	&	-0.099	&	0.608	&

\\ \hline
\\
\end{tabular}

\begin{minipage}{14.1cm}
Note: Mean, Standard deviation (SD), min, max, skewness (SKEW), kurtosis (KURT),  and Kendall's $\tau$ were calculated from  weekly percentage returns  of stock indices and 10-year government  bond (10 Yr Gov B) prices from December 2006 to December 2017.  
\end{minipage}
\end{center}

\end{table}

We estimate Equation (\ref{copulad1}) using the weekly data of stock price indices and 10-year government bond yields downloaded from 
\textbf{investing.com}.

The sample period from  December 2006 to December 2017 includes  the EU sovereign debt crisis. The EU crisis emerged when the underreported Greek government debt became evident in November 2009. The crisis started to calm down when the Outright Monetary Transaction was announced by the European Central Bank in 2012. Ireland and Spain exited from the bailout program at the end of 2013, though some countries had turbulent financial markets. 

\subsection{Model and Estimation}

Estimation is performed as follows:
\begin{enumerate}

\item Calculate stock and bond price returns by
\begin{equation}\label{returns}
\textrm{Stock Price Returns: } R_{Stock}(t) = (\log{ P_{Stock}(t)} - \log{P_{Stock}(t-1)})\times 100, 
\end{equation}
\begin{equation}\label{returns2}
\textrm{Bond Price Returns: }R_{Bond}(t) = (\log P_{Bond}(t) - \log P_{Bond}(t-1))\times 100,
\end{equation}
where $P_{Stock}(t)$ is the stock price, $P_Y(t)$ is the 10-year bond yield, and $P_{Bond}(t)$ is the bond price constructed by 
\begin{equation}\label{bond}
\textrm{Bond Price: }P_{Bond}(t)=\frac{1}{(1+P_Y(t))^{10}}.
\end{equation}
Table \ref{summary} shows the summary statistics of  the data.

\item AR(p)-GARCH(1,1) model with Student's $t$-distribution with degrees of freedom $\lambda$ is fitted to the marginal distribution of the stock and bond returns. The model is formalized as follows: 
\begin{eqnarray}\label{garch}
\begin{split}
R_j(t) &= \mu_j+\sum_{i=1}^{p_j}\gamma_{i,j}R_j(t-i)+e_j(t), \,e_j(t)=\sigma_j\epsilon_j(t),\,\\
 \epsilon_j(t)&\sim t(\lambda_j)\,\, \textrm{independently},\\
\sigma^2_j(t)& = \alpha_{j0}+\alpha_{j1}(t-1)+\beta_{j1}\sigma^2_j(t-1),\,\,j=\textrm{Stock, \, Bond}
\end{split}
\end{eqnarray}

Standardized residuals  $\hat\epsilon_{Stock}(t)$ and $\hat\epsilon_{Bond}(t)$ estimate  $\epsilon_{Stock}(t)$ and $\epsilon_{Bond}(t)$, respectively, using the estimated AR-GARCH model (\ref{garch}). Their estimates are summarized in Table \ref{AR}.

\item
The GARCH residuals are transformed to uniformly distributed variables by $X=F_t(\hat\epsilon_{Stock}, \hat\lambda_{Stock})$ and $Y=F_t(\hat\epsilon_{Bond}, \hat\lambda_{Bond})$, where $F_t(\cdot, \hat\lambda_{Stock})$ and $F_t(\cdot , \hat\lambda_{Bond})$ are $t$-distribution functions with estimated degrees of freedom of $\hat\lambda_{Stock} $ and $\hat\lambda_{Bond} $, respectively.
\item 
The likelihood function is defined by  (\ref{copulad1}), 
under the assumption that $X=F_t(\hat\epsilon_{Stock}, \hat\lambda_{Stock})$ and $Y=F_t(\hat\epsilon_{Bond}, \hat\lambda_{Bond})$ follow split normal copula distribution.
\item 
The parameters of the split normal copula are estimated using 100-week rolling MLE by maximizing the log of (\ref{copulad1}) using the grid search method with grid width 0.01. The result is presented in Figure \ref{6countries}.

\end{enumerate}

\begin{table}

\begin{center}
\caption{Summary Statistics of AR($p$)-GARCH(1,1) Estimation}\label{table:garch}

\begin{tabular}{lllllllll}Country&Assets&AR($p$)&\multicolumn{3}{c}{GARCH(1,1)}&&
 \\ &&$p$&$\alpha_0$&$\alpha_1$&$\beta_1$&$\lambda$

\\ \hline \hline

       \multirow{4}{*}{Greece }&Stock 		&	0	
       
     &		0.592		&		0.181		&		0.810		&		16.260		\\

&	&

&	( 0.338 )	&	( 0.050 )	&	( 0.049 )	&	( 8.993 )	\\
&Bond&	0	

&		0.138		&		0.216		&		0.783		&		3.606		\\
&	&

&	( 0.039 )	&	( 0.033 )	&	( 0.030 )	&	( 0.365 )	\\

\hline

\multirow{4}{*}{ Ireland} &  Stock
&	0	

&		0.186		&		0.094		&		0.896		&		5.705		\\

&	&

&	( 0.120 )	&	( 0.026 )	&	( 0.026 )	&	( 1.256 )	\\
&Bond	&	0	

&		0.136		&		0.221		&		0.775		&		3.419		\\
&	&

&	( 0.059 )	&	( 0.075 )	&	( 0.051 )	&	( 0.558 )	\\

\hline
\multirow{4}{*}{Italy }&  Stock&		0

&		0.248		&		0.112		&		0.871		&		8.861		\\
&		&		

&	( 0.157 )	&	( 0.033 )	&	( 0.036 )	&	( 2.418 )	\\
&	Bond	&	0	

&		0.049		&		0.090		&		0.883		&		6.387		\\
&		&

&	( 0.055 )	&	( 0.033 )	&	( 0.028 )	&	( 1.563 )	\\

 \hline
\multirow{4}{*}{Portugal} & Stock
	&	0	

&		0.413		&		0.200		&		0.773		&		4.814		\\
&		&

&	( 0.332 )	&	( 0.072 )	&	( 0.089 )	&	( 0.990 )	\\
&	Bond&	0	

&		0.041		&		0.123		&		0.876		&		4.806		\\
&		&

&	( 0.037 )	&	( 0.026 )	&	( 0.024 )	&	( 0.887 )	\\

 \hline
\multirow{4}{*}{Spain}&  Stock
	&	1

&		0.349		&		0.098		&		0.872		&		7.516		\\
&		&	
&	( 0.230 )	&	( 0.033 )	&	( 0.045 )	&	( 1.881 )	\\

&	Bond	&	1

&		0.033		&		0.101		&		0.886		&		6.681		\\
&		&

&	( 0.042 )	&	( 0.039 )	&	( 0.042 )	&	( 1.687 )	\\

\hline
\multirow{4}{*}{Germany}&Stock
	&	1

&		0.440		&		0.121		&		0.828		&		6.724		\\
&		&	

&	( 0.229 )	&	( 0.046 )	&	( 0.061 )	&	( 1.531 )	\\
&	Bond	&	1

&		0.015		&		0.066		&		0.920		&		35.556		\\
&		&		

&	( 0.097 )	&	( 0.110 )	&	( 0.125 )	&	( 54.324 )	\\

\hline
\\
\end{tabular}
\label{AR}
\begin{minipage}{13.1cm}
Note: The order of AP($p$) is chosen by minimizing AIC. Standard errors are shown in parentheses.
\end{minipage}\end{center}
\end{table}

Figure \ref{6countries} shows the estimates of  upper and lower tail correlation coefficients obtained by the rolling MLE with 100-week rolling window using grid search with grid width  0.01.
\begin{figure}[H]
  \begin{center}
 \includegraphics[clip,scale=1.0]{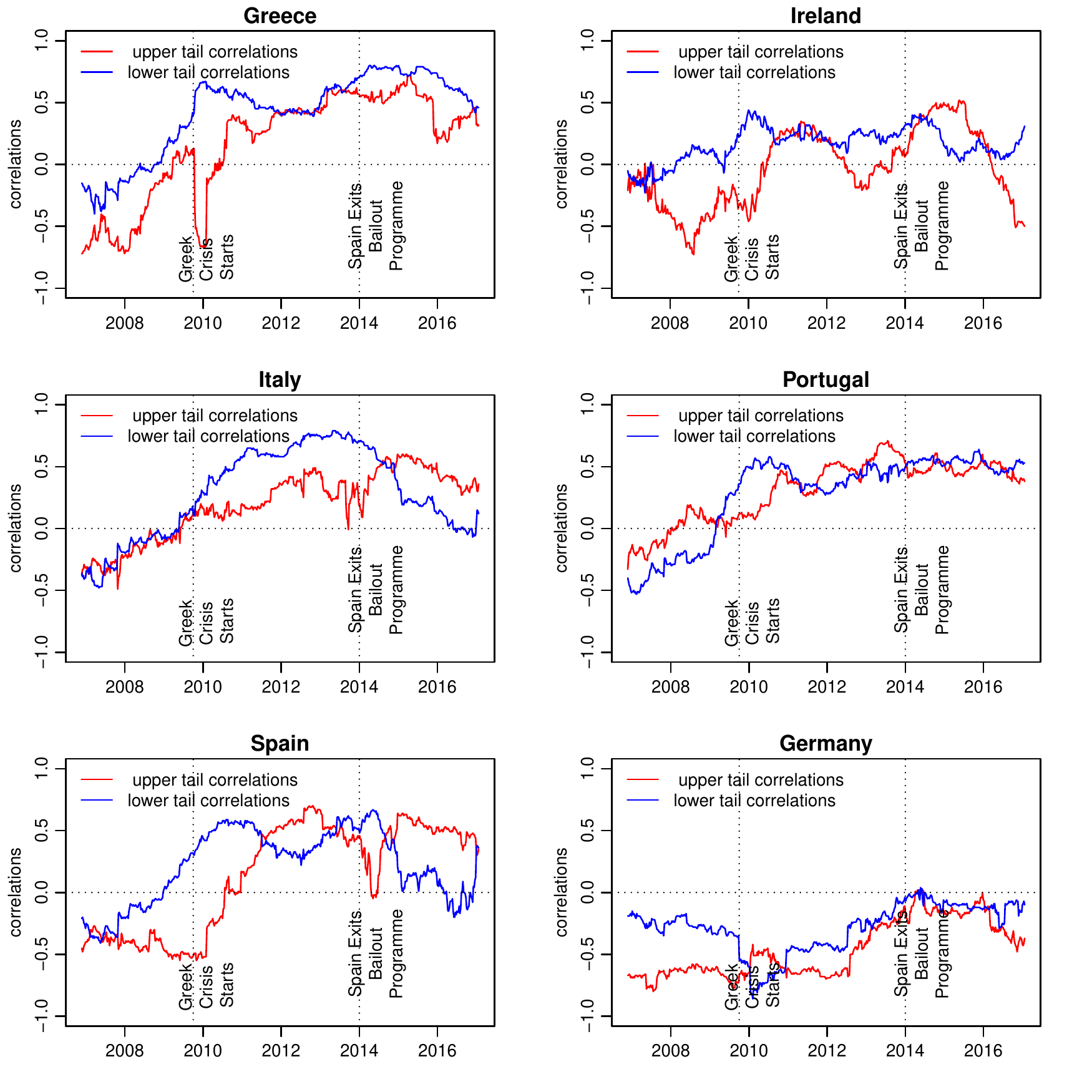}

\begin{minipage}{13.1cm}
  \caption{Upper and Lower Tail Correlation of Stock-Bond distribution. The lines show the 100-week rolling window maximum likelihood estimates of the upper and lower tail correlation coefficients of the split normal copula model. \label{6countries}
}
\end{minipage}
  \end{center}

\end{figure}
 The estimation result illustrated in Figure \ref{6countries}   is summarized as follows:
Before the EU crisis, the signs of stock-bond correlation of all the countries were negative or near zero. 
Shortly after the EU crisis emerged at the end of 2009 the EU periphery countries had  large positive lower tail correlations and  near-zero or negative upper tail correlations. , namely  when the price of one asset decreases, the price of another asset also decreases, whereas when the price of one asset increases, the price of another asset does not always increase. It is suspected that this asymmetry is the result of  capital flight from these countries.
The stock-bond correlations of the EU periphery countries became positive in both tails  some years after the crisis started, which suggests that the government bonds of these countries were not safe assets.
In contrast, Germany had negative or near-zero correlations throughout the sample period, unlike the EU periphery countries. The stock-bond correlations of the Netherlands and France, which are core EU countries, had a similar pattern to that of Germany, though they are not reported in this paper.



\subsection{Asymmetry and Correlation  Reversal}
This subsection summarizes asymmetry and correlation reversal of the stock and bond yields of Spain in detail. The other EU peripheral countries shows essentially similar patterns. 
Figures \ref{fig:contour1} and \ref{fig:contour2}  exhibit  the four patterns of asymmetry and  correlation reversal of the split normal copula   in  financial crisis using the estimated parameter values of Spain. 
The first row of Figures \ref{fig:contour1} and \ref{fig:contour2}  shows contour plots of the split normal copula density (\ref{copulad1}) transformed so as to have standard normal marginals.  The second row shows the scatter plot of  normalized residuals of  estimated GARCH model for the rolling window periods used in the first and second rows.

In the first column of Figure \ref{fig:contour1}  the  upper and lower tail correlations were both negative before the financial crisis. In the second column  the  lower tail correlation became positive and the  upper tail correlation was still negative. This asymmetry implies that  the price of  an asset  decreases with high probability when the price of the other asset decreases; when the  price of an asset increases, the  price  of the other asset does not  always increase. In the first column of Figure \ref{fig:contour2}   upper and lower tail correlations were both positive in the middle of the crisis, and the upper tail correlation was positive and the   lower tail correlation was negative in the second column. 

Table \ref{spain}  summarizes the result of one-sided hypothetical testing for the null hypothesis  of zero  correlation coefficient. 

\begin{figure}[H]
  \begin{center}
    \includegraphics[width=15cm]{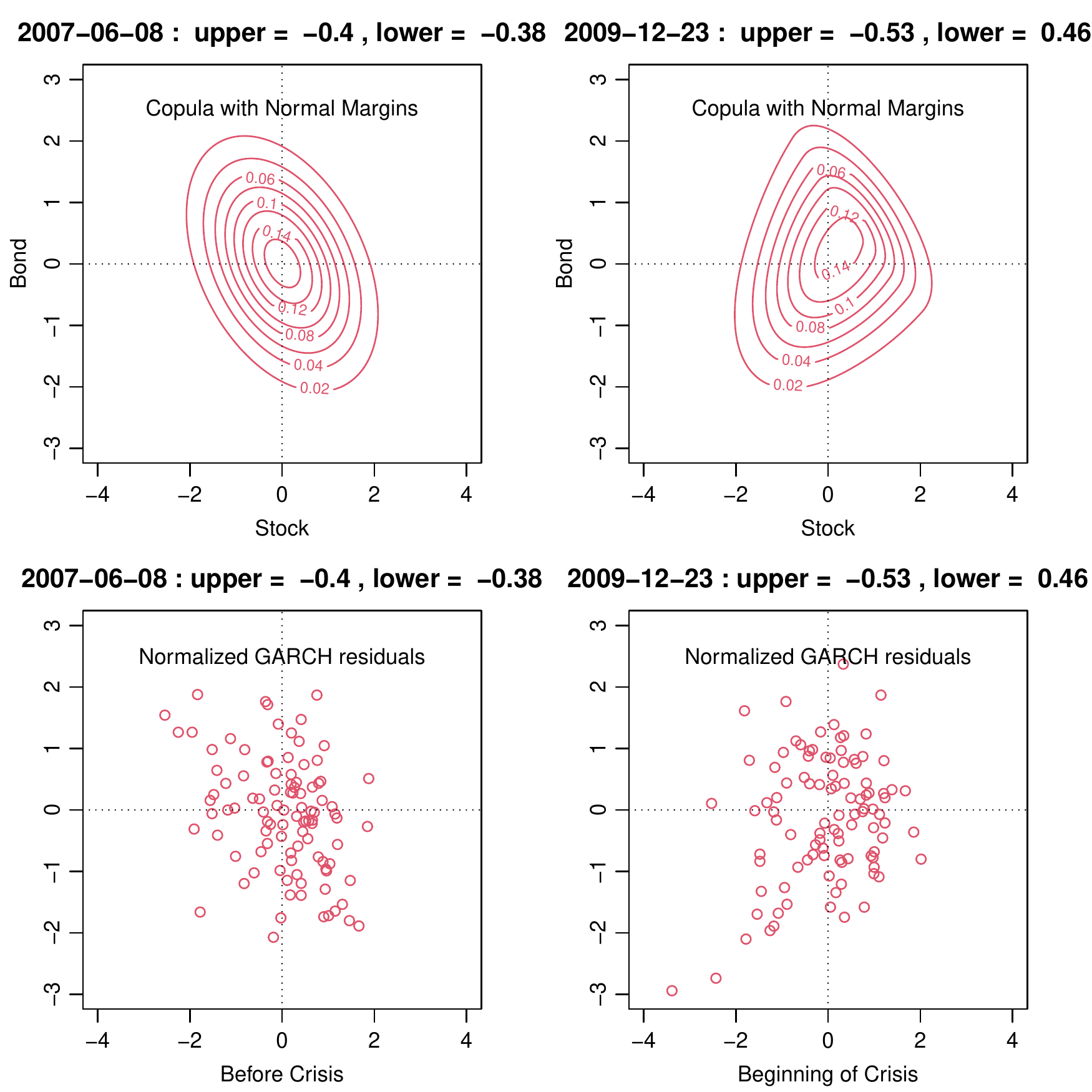}

  \caption{The first row shows contours of split normal copula where the marginal variables are transformed so as to have  normal marginal distribution. The second row shows the scatter plot of the normalized GARCH residuals of   Stock and Bond returns . All the plots use the data and correlation coefficient estimates of Spain for  a subsample of 100 weeks obtained by the maximum likelihood method.}   
 \label{fig:contour1}
  \end{center}

\end{figure}

\newpage

\begin{figure}[H]
  \begin{center}
   \includegraphics[width=15cm]{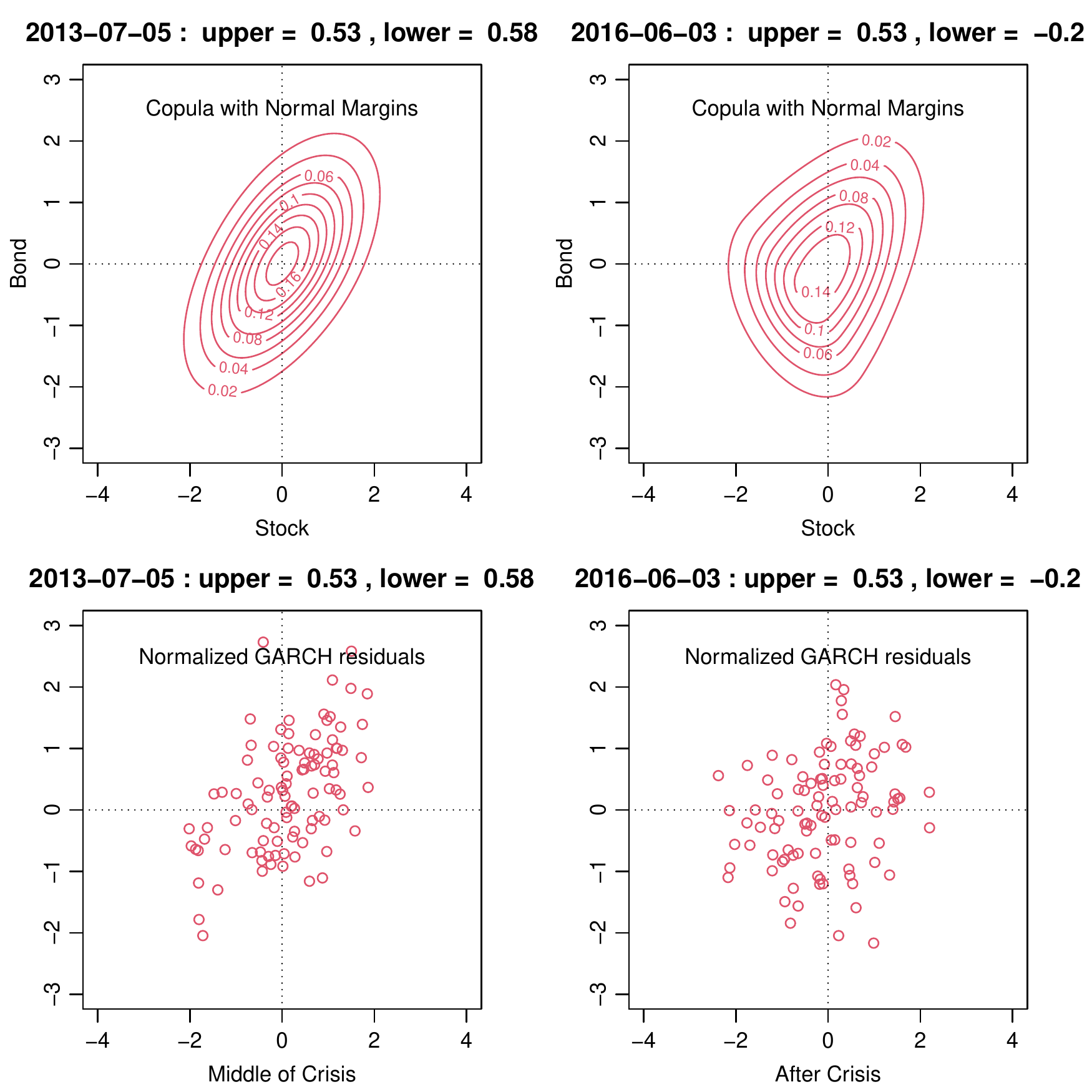}

  \caption{Contour plots of the split normal copula transformed so as to have standard normal marginal in the first row. The second row shows the scatter plot of the normalized GARCH residuals of   Stock ($x$-axis) and Bond ($y$-axis) returns.  All the plots use the data and correlation coefficient estimates of Spain for  a subsample of 100 weeks using the maximum likelihood method.}   \label{fig:contour2}
 
  \end{center}

\end{figure}

\begin{table}

\begin{center}
\caption{Estimated upper and lower tail corrleation of Spain and their statistical significance }
\begin{tabular}{ccccccccccc}
\\
\\Period&\multicolumn{2}{c}{Upper Tail Correlation $\rho_U$}& \multicolumn{2}{c}{LowerTail Correlation $\rho_L$ } \\
Center Date	&	  			Estimate&	Statistical Significance &Estimate&	Statistical Significance \\\hline
2007-06-08		&	-0.40	&Negative at 0.05 	&		-0.38&	Negative at 0.05 \\
2009-12-23	&	 -0.53		&Negative at 0.10 &		0.46&	Positive at 0.05 \\
2013-07-05	&	0.53		&Positive at 0.05 &		0.58&	Positive at 0.05\\
2016-06-03	&	 0.53	&Positive at 0.05 	&		-0.20&	Not Significant \\
\hline
\\
\end{tabular}
\label{spain}
\label{cv}
Note: The upper  and lower tail correlation coefficients were estimated from  100-week samples.  
Statistical significance is based upon one-sided tests using   Table  \ref{cv}.

\end{center}

\end{table}
 
\section{Conclusion}
   
This study proposed a novel two-parameter copula that can show asymmetry and  reversal of  correlation coefficients in the upper and lower tail. 
This method was applied to the  stock-bond distribution of  the periphery countries in the EU sovereign debt crisis and revealed that  the sign of correlation coefficient changed  from negative to positive, and that  asymmetry, namely  positive lower tail correlation and negative or near-zero upper tail correlation, emerged at the beginning of EU crisis. This result suggests capital flight from these countries. 
 In contrast, Germany had negative stock-bond correlation throughout the EU sovereign debt crisis.

\section*{References}

\parindent=-15pt
\indent

Cai, Z. and X. Wang (2014) Selection of Mixed Copula Model via Penalized Likelihood, \textit{Journal of the American Statistical Association}, 109, 788-801.

Christoffersen, P., V. Errunza, and K. Jacobs (2012) Is the Potential for International Diversification Disappearing? A Dynamic Copula Approach, 
\textit{Review of Financial Studies}, vol. 25, no. 12, 3711-3751.

Dufour, A., A. Stancu, and S. Varotto (2017) The Equity-like Behaviour of Sovereign Bonds,\textit{Journal of International Financial Markets, Institutions \& Money}, vol. 48, 25-46.

Engle, R. (2002)Dynamic Conditional Correlation:
A Simple Class of Multivariate Generalized Autoregressive Conditional
Heteroskedasticity Models, \textit{Journal of Business and Economic Statistics}, vol. 20, no. 3, 339-350.

Geweke, J. (1989) Bayesian Inference in Econometric Models Using Monte Carlo Integration,
\textit{Econometrica}, vol. 57, no. 6, 1317-1339


Li, F. and Y. Kang (2018) Improving Forecasting Performance Using
Covariate-Dependent Copula Models, \textit{International Journal of Forecasting,}  34, 456-476.

Liu, G., Long, W., Zhang, X., and Li, Q. (2019) Detecting Financial Data Dependence Structure By Averaging Mixture Copulas
, Econometric Theory, 35(4), 777-815. 

Nelson, R. B. (2006) \textit{An Introduction to Copulas-Springer Series in Statistics}, Springer, New York, USA.

Ohmi, H. and T. Okimoto (2016) Trends in Stock-Bond Correlations, \textit{Applied Economics,}  48, 536-552.

Okimoto, T. (2008) New Evidence of Asymmetric Dependence Structures in
International Equity Markets, \textit{Journal of Financial and Quantitative Analysis} vol. 43, no. 3, 787-816.

Patton, A. J. (2006) Modelling Asymmetric Exchange Rate Dependence,
\textit{ International Economic Review}, vol. 47, no. 2, 525-555.

Sklar, A. (1959) Fonctions de r\'epartition \`{a} \textit{n} dimensions et leurs marges, \textit{Publications de L'Institut de Statistique de L'Universite de Paris}, vol. 8, 229-231.

Villani, M. and R. Larsson (2006) The Multivariate Split Normal Distribution and Asymmetric Principal Components Analysis, \textit{Communications in Statistics - Theory and Methods}, vol. 35, no. 6, 1123-1140.

Yang, B., Z. Cai, C. Hafner, and G. Liu (2018) Trending Mixture Copula Models with Copula Selection, IRTG 1792 Discussion Paper 2018-057.

Yoshiba, T. (2013) Risk Aggregation by a Copula with a Stressed Condition', Bank of Japan Working Paper Series 13-E-12, Bank of Japan.

\newpage

\newpage

\end{document}